# GRAIN ALIGNMENT IN THE TAURUS DARK CLOUD

P. A. Gerakines [1,2] and D. C. B. Whittet [2]

[1] Leiden Observatory, P.O. Box 9513, 2300 RA Leiden, The Netherlands
[2] Department of Physics, Applied Physics & Astronomy, Rensselaer Polytechnic Institute, Troy, New York 12180

# ABSTRACT

Variations in polarization efficiency ($p/A$) of interstellar grains as a function of environment place vital constraints on models for the mechanism of their alignment. We use polarimetry of background field stars to investigate alignment in the Taurus dark cloud as a function of optical depth for extinctions of $A_K < 2.5$ ($A_V < 25$). Results show a strong systematic trend in polarization efficiency with extinction. We discuss the possibility of a relationship between $p/A$ and icy grain mantle growth, since mantles are detected in Taurus only for extinctions where $p/A$ is lowest. A simple density model of alignment by paramagnetic alignment (Davis & Greenstein 1951) predicts that the magnetic field strength ($B$) scales with density to the power $\kappa = 0.28$.

*Subject headings:* dust, extinction — polarization — ISM: magnetic fields — ISM: individual (Taurus Cloud)

# 1 INTRODUCTION

The light from stars obscured by interstellar dust is almost always partially plane polarized at visible wavelengths, a phenomenon attributed to extinction by aspherical dust grains which are aligned to some degree by the interstellar magnetic field (see Whittet 1995 for a review). It is widely agreed that alignment is the result of an interaction between the rotational dynamics of the grains and the ambient magnetic field, as originally proposed by Davis & Greenstein (1951). Paramagnetic relaxation results in the grains tending to become oriented with their angular momenta parallel (and hence their long axes perpendicular) to the magnetic field lines. Although this mechanism is qualitatively highly successful, a detailed quantitative theory of magnetic alignment that is consistent with all existing observational constraints is yet to be formulated. The efficiency of alignment depends on both the properties of the dust grains themselves and on the environment in which they exist (e.g. Jones & Spitzer 1967; Purcell 1979; Mathis 1986; DeGraff et al. 1995). Polarization studies, in conjunction with measurements of other physical and chemical characteristics of dense clouds, may thus be used as observational tools for constraining the alignment mechanism.

The observed degree of linear polarization ($p$) correlates positively with the magnitude of extinction ($A$) due to dust in the line of sight, but with scatter much greater than can be accounted for by observational errors alone (e.g. Serkowski et al. 1975). The polarization efficiency of the interstellar medium (ISM), represented by the ratio $p/A$ at some wavelength, is highly non-uniform. In general, $p/A$ falls below some upper boundary[3] representing optimum efficiency. Scatter in $p/A$ below this optimum value might arise due to a number of factors. In a uniform medium with a uniform magnetic field, $p/A$ varies systematically with the angle between the line of sight and the field direction, and is maximum when this angle is 90°. If the magnetic field is inhomogeneous, the direction of alignment may change along a given line of sight, leading to a reduction in net polarization. Similar depolarization effects may be observed if two or more clouds with different field orientations exist within the same line of sight, and if the spectral dependence of polarization also changes from cloud to cloud, the degree of depolarization becomes a function of wavelength. Of greater interest are variations in $p/A$ induced by changes in the efficiency of alignment as a function of physical conditions. In this paper, we attempt to isolate the effect of physical environment on polarization efficiency in order to improve our understanding of the alignment mechanism.

The effects of viewing geometry and depolarization may be minimized by appropriate choice of sample. Our approach is to select data for stars obscured by an individual cloud displaying a relatively uniform magnetic field structure. The Taurus dark cloud is an excellent candidate. As it is relatively nearby ($\sim$ 140 pc, Elias 1978) and $\sim$ 20° from the Galactic plane, the extinctions (and polarizations) of stars in this area of the sky arise almost entirely within the cloud itself (Straizys & Meistas 1980; Kenyon et al. 1994). The position angle of polarization is independent of wavelength in 15 out of 21 stars studied by Whittet et al. (1992), consistent with a general absence of discrete cloud components with different wavelength dependencies of polarization. The distribution of polarization vectors in the plane of the sky is fairly uniform over the entire cloud (Moneti et al. 1984; Tamura et al. 1987; Goodman et al. 1992). Although small-scale regions of magnetic complexity associated with individual condensations undoubtedly occur, the macroscopic field (as sampled by field stars) does appear to be generally lacking in major inhomogeneities. We therefore surmise that the degree of grain alignment is the primary factor determining the polarization efficiency, $p/A$, in the Taurus dark cloud.

---

[3] In the diffuse ISM, a limiting value given by $p_{\max}/A_V < 3.0$ % mag$^{-1}$ is generally assumed (e.g. Serkowski et al. 1975), where $p_{\max}$ is the peak percentage polarization in the Serkowski formulation, $A_V$ is the visual extinction, and $A_V = 3E_{B-V}$ has been assumed. Whittet et al. (1994) recently found values of $p_{\max}/A_V$ up to $\sim$ 4.5 % mag$^{-1}$ in the Chamaeleon I dark cloud.

A further advantage of this choice is the fact that the Taurus cloud has been well studied in the previous literature and consequently its physical and chemical properties are reasonably well constrained. Studies of the Taurus-Auriga cloud complex show that it possesses a filamentary structure (e.g. Kleiner & Dickman 1984, Cernicharo et al. 1985, Wouterloot & Habing 1985), with small dense cores observed in line emission of molecules such as $NH_3$ (e.g. Myers & Benson 1983) and CS (Zhou et al. 1989). A complex chemistry is evident in these cores, with such varied molecules as $C_3N$ (Friberg et al. 1980), HNCO (Brown 1981), $CH_3CCH$ (Irvine et al. 1981), and aromatic hydrocarbons (Hanner et al. 1994) detected in the gas phase. They have typical masses of a few $M_\odot$, gas temperatures of $\sim 10$ K, and densities $\sim 10^4$ cm$^{-3}$ (e.g. Wilson & Minn 1977), and are thought to be active sites of star formation (e.g. Lada et al. 1993), associated with T Tauri stars (Myers & Benson 1983) and low luminosity IRAS sources (Beichman et al. 1986). The magnetic field strengths ($B$) in the dense cores are estimated to be $\sim 60 - 300\mu$G, assuming that the field within the cloud scales as $B = B_0(n/n_0)^\kappa$ (Mouschovias 1978), where $\frac{1}{3} < \kappa < \frac{1}{2}$, $n_0 = 1$ cm$^{-3}$ and $B_0 = 3\mu$G. The range given for $\kappa$ is governed by observational constraints and theoretical calculations for the lower and upper limits, respectively, the latter more closely defined than the former (see discussion in Mouschovias 1978).

Dust grains in the outer layers of the cloud ($A_V < 3$) appear to have optical properties similar to those in the diffuse ISM (Vrba & Rydgren 1984; Kenyon et al. 1994). In lines of sight that sample regions of higher density, infrared spectroscopy of the 3.0$\mu$m water-ice feature demonstrates the presence of icy mantles on the grains (Whittet et al. 1988). The transition from bare to mantled grains appears to occur at a rather well-defined value of $A_V = 3.3 \pm 0.1$ (the threshold extinction). For lines of sight with extinctions higher than this threshold value, the optical depth of the 3.0$\mu$m feature (and hence the column density of water molecules on the grains) increases linearly with $A_V$. The feature of solid CO at 4.67$\mu$m demonstrates a similar behavior, with a higher threshold extinction of $A_V \sim 6$ mag (Whittet et al. 1989, Chiar et al. 1995). Mantle growth represents a total change in grain surface properties, upon which some alignment mechanisms critically depend, and it is therefore of interest to compare $p/A$ at extinctions above and below these thresholds.

In the following sections, we investigate the correlation between $p/A$ and $A$ in the near infrared ($K$ band; $\lambda_0 = 2.2\mu$m) for field stars background to the Taurus dark cloud. We make use of previous measurements of polarization together with a refinement of values for extinction from the literature. In § 2 we present our data compilation and results. Implications are discussed in § 3, and we summarize our conclusions in § 4.

## 2 THE DATA

As compilations of visual polarization are heavily biased to low-extinction objects, observations in the $K$ passband offer a better opportunity to construct a homogeneous set of data covering stars with a wide range of extinctions. Values of $p_K$ listed in Table 1 are derived from Moneti et al. (1984), Tamura et al. (1987), Goodman et al. (1992) and Whittet et al. (1992, 1995). Data were rejected in cases where the fractional error ($\sigma_{p_K}/p_K$) is found to be greater than 20%.

Measurements of $p_K$ and $\sigma_{p_K}$ are explicitly given in three references (Whittet et al. 1995, Tamura et al. 1987 and Goodman et al. 1992). Polarimetric data from Moneti et al. (1984) are listed in the $V$ passband and converted using an average ratio of $p_K/p_V = 0.20$, which was calculated using Taurus objects from Whittet et al. (1992). Note that the $p_K$ values of Whittet et al. (1992) are measured in a non-standard passband centered at 2.04$\mu$m. In these cases, our $p_K$ values are determined using a power law extrapolation of the listed values of $p$ at 1.64 and 2.04$\mu$m (Whittet et al. 1992, Table 3), except in four cases where the value at 2.04$\mu$m was unknown (see Table 1). The Serkowski formula

$$p/p_{\max} = \exp[-K \ln^2(\lambda_{\max}/\lambda)] \tag{1}$$

(Serkowski 1973; Coyne et al. 1974; Serkowski et al. 1975) was used to derive a value of $p_K$ for these stars, with the parameters listed in Table 4 of Whittet et al. (1992).

Corresponding extinction data ($A_K$) are also listed in Table 1. In the cases where the spectral type is known, visual and infrared photometry from the literature (see Table 1 for references) are used to derive infrared extinctions ($A_K$) assuming a standard extinction law: $A_K = 0.1 A_V$, where $A_V = 1.1 E_{V-K}$ (Whittet & van Breda 1978).

In the cases where the spectral types are not known (as in Tamura et al. 1987 and Goodman et al. 1992), a spectral type of K5III and an intrinsic color of $(J-K)_0 = 0.95 \pm 0.4$ are assumed. This spectral type corresponds to the mean of the Elias (1978) set of objects, where the error in intrinsic color reflects the range in $(J-K)_0$ from types K0III through M8III (the span of the Elias 1978 set). Intrinsic colors are taken from Bessell & Brett (1988).

For stars with the highest extinctions, only photometry in the near-infrared passbands ($J$, $H$, $K$) could be found in the literature. Hence, we use an empirical formula derived by Whittet (1992) for extinction based upon infrared color excess, $E_{J-K}$:

$$A_K = 0.1 \, r \, E_{J-K} \, , \tag{2}$$

where $r$ may be determined from the total to selective extinction ratio ($R_V$) by the formula

$$r = \frac{2.332}{0.778 - 1.164 R_V^{-1}} \, . \tag{3}$$

A total of 14 Taurus sources whose photometric magnitudes in the $B$, $V$ and $K$ passbands are known (Whittet et al. 1995) give a mean value of $R_V \approx 3.1$, leading to $r = 5.8$ from equation (3). Thus, we use this value in equation (2) in our calculations of $A_K$.

In the case of four stars from Whittet et al. (1995), we combine known values of $\lambda_{\max}$ and $E_{B-V}$ with the relation $R_V = 5.6\,\lambda_{\max}$ (Whittet & van Breda 1978) to derive $A_K = 0.56\,\lambda_{\max}\,E_{B-V}$ (see Table 1).

In Figure 1, we plot the resulting values of $p_K/A_K$ as a function of $A_K$ for the stars listed in Table 1.

## 3  DISCUSSION

Figure 1 displays a monotonically decreasing relationship between alignment efficiency ($p/A$) and extinction ($A$), indicating that the linear polarization of radiation due to aligned dust grains in the Taurus dark cloud is produced more efficiently in regions of low extinction than in the more obscured lines of sight. Several physical processes may contribute to this trend in $p/A$. Similar decreases in polarization efficiency in the outer regions of dark clouds have been previously observed in $\rho$ Ophiuchi (Vrba et al. 1976,1993), R Coronae Australis (Vrba et al. 1981), and Chamaeleon I (McGregor et al. 1994; Whittet et al. 1994). It may be worth noting that similar declines with extinction have been observed for the diffuse band (DIB) production efficiencies ($W_\lambda/E_{B-V}$) in $\rho$ Oph by Snow & Cowen (1974) and in Taurus by Adamson et al. (1992), although no direct relationship between DIB's and aligned interstellar dust grains has ever been established.

If one considers only the effects of density ($n$) on classical DG alignment, it is possible to derive the density dependence of the cloud's magnetic field strength, $B$ (following the method of Vrba et al. 1981). For grains of equal radius,

$$P/A_{\mathrm{pg}} = 2.07 F \cos^2 \nu \frac{(\sigma_a/\sigma_b) - 1}{(\sigma_a/\sigma_b) + 2} \;, \qquad (4)$$

(Davis & Greenstein 1951), where $P$ is the polarization in magnitudes, $A_{\mathrm{pg}}$ is the photographic extinction ($A_{\mathrm{pg}} \approx A_B$), $F$ is the distribution parameter of grain orientations, $(\frac{\pi}{2} - \nu)$ is the angle between the direction of the magnetic field ($\hat{B}$) and the line of sight, and $\sigma_a$ and $\sigma_b$ are the extinction coefficients parallel and perpendicular to the long axes of the grains, respectively. Assuming that $\hat{B}$ is perpendicular to the line of sight, then $\cos^2 \nu = 1$. If we take $\sigma_b/\sigma_a = b/a$ (Davis & Greenstein 1951, Vrba et al. 1981) and a typical interstellar grain axial ratio of $b/a = 0.2$ (Aanestad & Purcell 1973), equation (4) reduces to $P/A_{\mathrm{pg}} = 1.2\,F$. Conversion of $A_{\mathrm{pg}}$ to $A_K$ and $P$ to $p$ (in %) gives $p/A_K = 730\,F\%$ mag$^{-1}$. From Jones & Spitzer (1967),

$$F = \frac{\chi'' B^2}{75 a \omega n}\left(\frac{2\pi}{mkT}\right)^{1/2}(\gamma - 1)\left(1 - \frac{T_{\mathrm{gr}}}{T}\right)\;, \qquad (5)$$

where the imaginary part of the grain's magnetic susceptibility ($\chi''$) and its rotational velocity ($\omega$) are related as $\chi''/\omega = 2.6 \times 10^{-12} T_{\mathrm{gr}}^{-1}$ for reasonable grain materials and values of $\omega$ (Purcell 1969). $T_{\mathrm{gr}}$ is the grain temperature, $T$ the surrounding gas temperature, $B$ the strength of the magnetic field, $a$ the grain size, and $\gamma$ the ratio of the moments of inertia about the short and long axes, respectively, $\gamma = \frac{1}{2}[(a/b)^2 + 1]$ (Davis & Greenstein 1951). Taking $T = 50\mathrm{K}$, $T_{\mathrm{gr}} = 10\mathrm{K}$ as typical for dense clouds outside of dense cores and combining equations (4) and (5), we find

$$p/A_K = \frac{5.7 \times 10^8 B^2}{an} \% \text{ mag}^{-1}. \qquad (6)$$

Assuming a density dependence for the magnetic field strength of $B = B_0(n/n_0)^\kappa$ (Mouschovias 1978) and using the mean interstellar values of $B_0 = 3\ \mu\mathrm{G}$ and $n_0 = 1\mathrm{cm}^{-3}$, we derive

$$p/A_K = \frac{5.1 \times 10^{-3}}{a} n^{2\kappa - 1} \% \text{ mag}^{-1}. \qquad (7)$$

Since $n = N/L$ (where $N$ is the column density of the cloud and $L$ is the length of the line of sight through it) and it is known from observations that $A \propto N(\mathrm{H})$, the power law dependence on $n$ in equation (7) also holds for $A$. Thus, $p/A_K \propto A_K^{2\kappa - 1}$.

A least-squares power law fit to all data points in Table 1 yields

$$p_K/A_K = 1.38\ A_K^{-0.53} \% \text{ mag}^{-1}. \qquad (8)$$

Relating this to equation (7), we derive a scaling factor for $B$ of $\kappa = 0.24$ for the Taurus dark cloud.

This model considers the scaling of $B$ with $n$, ignoring the effect of grain growth. The relative enhancements of the ratio of total to selective extinction ($R_V$) and the wavelength of maximum polarization ($\lambda_{\max}$) toward dense regions of dark clouds suggest that grain growth by mantle condensation or grain-grain coagulation occurs there, producing grains which are more spherical and thus reducing the efficiency with which they may be aligned (Carrasco et al. 1973; Whittet & van Breda 1975; Whittet & Blades 1980). Models of polarization due to alignment by the DG mechanism show that in fact grains which are less elongated do have lower values of $p/A$ (Purcell & Spitzer 1971; Aannestad & Purcell 1973). On the basis of the variation in $\lambda_{\max}$ with $A_V$ in the $\rho$ Oph cloud, Vrba et al. (1993) deduce that the effect of grain growth may be described as $a = 1.18\ a_i\ A_V^{1/12}$, where $a_i$ is the initial value of $a$. This introduces an additional factor of $n^{-1/12}$ to equation (7), and hence, taking this into account, $\kappa = 0.28$. This value is somewhat below the lower limit of $\frac{1}{3}$ given by Mouschovias (1978), but within a reasonable range considering the assignment of this limit, and also consistent with the results of Tamura et al. (1987) and Goodman et al. (1992).

Stronger declines in $p/A$ with extinction may occur due to several different processes within dense interstellar clouds. Grain surfaces are active sites for the formation of $H_2$. A hydrogen atom colliding with a dust grain has a high probability of sticking, followed by migration over the surface until finally combining with another hydrogen atom (e.g. Tielens & Allamandola 1987). The energy of this formation process may be transferred in part to kinetic energy, ejecting the molecule from the surface and donating angular momentum to the grain. Molecular formation is likely to occur at preferential sites on the grain surface due to defects or impurities in the surface structure (Hollenbach & Salpeter 1971), and if the number of active sites are small, then their distribution over the surface will determine the spin properties of the grain (Purcell 1975, 1979). A series of recombination events occurring at these preferential sites could then lead to grain angular velocities that are a factor of $\sim 10^5$ higher than those predicted by random elastic gas-grain collisions. This suprathermal spin effect would lead to very efficient alignment of interstellar grains, but the physical processes required may not be available in all grain environments.

Undoubtedly, the largest influence on the suprathermal spin mechanism is that the abundance of atomic hydrogen available with which to form $H_2$ falls rapidly with increasing cloud density (Savage et al. 1977). Moreover, declines in $p/A$ could be expected solely on the basis of $H_2O$ ice mantle growth since hydrogen adsorption sites on an $H_2O$ surface are weak physisorption sites, as opposed to the stronger sites obtainable on a bare silicate or carbonaceous surface. Hence, the time a hydrogen atom spends on an $H_2O$ surface (and thus the probability that it will combine with a second such adsorbed H atom) will be smaller than that on a bare, unmantled grain (Tielens & Allamandola 1987, Williams 1993).

Mathis (1986) suggest that a small inclusion of superparamagnetic materials (e.g. metallic Fe, $Fe_3O_4$, etc) in interstellar grains could increase the imaginary part of their magnetic susceptibility ($\chi''$) by a factor of $10^6$ over ordinary materials. While superparamagnetic (SPM) alignment could efficiently align grains even in the observed galactic magnetic field, its environment dependence is not different from that of DG alignment.

## 4 CONCLUSIONS

In combining several data sets for the Taurus dark cloud for a large range in extinction ($A_K < 2.5$ mag, $A_V < 25$ mag), we have shown that the efficiency with which background starlight is polarized by dust ($p/A$) decreases strongly with the amount of obscuring material in the line of sight. As previously noted, the form of Fig 1 is a monotonic decrease in $p/A$ with $A$. We have shown that this behavior is consistent with the expected dependence of $B$ on $n$ for DG-type alignment. Changes in grain surface properties associated with the appearance of $H_2O$ mantles ($A_K > 0.3$) and CO mantles ($A_K > 0.6$) do not have an obvious effect on the alignment efficiency within the scatter of our data, in spite of the fact that grains with icy surfaces should possess different suprathermal spin properties than bare grain cores. The presence (or absence) of mantles may have little effect on the ability of a grain to align in the ambient magnetic field. This is consistent with the observed presence of polarization enhancement in the $3\mu$m ice feature (Hough et al. 1988, 1989) which demonstrates that grains mantled with water-ice do indeed align. Small but non-zero alignment efficiencies deep within dark clouds are a natural consequence of alignment by the DG mechanism.

**Table 1.** Spectral type, $A_K$, $p_K$, $\sigma_{p_K}$, and $p_K/A_K$ for the set of objects studied

| Star I.D. | Sp Type[a] | $A_K$ mag | $p_K$ % | $\sigma_{p_K}$ % | $p_K/A_K$ % mag$^{-1}$ | Ref |
|---|---|---|---|---|---|---|
| Elias 3 | K2III | 0.97 | 0.65 | 0.02 | 0.67 | 1,2 |
| Elias 4 | K0III | 0.12 | 0.22 | 0.02 | 1.83 | 1,3 |
| Elias 15 | M2III | 1.85 | 2.1 | 0.4 | 1.14 | 1,4 |
| Elias 16 | K1III | 2.72 | 2.6 | 0.5 | 0.96 | 1,4 |
| Elias 19 | M4III | 0.17 | 1.20 | 0.03 | 7.06 | 1,5 |
| Elias 29 | G9III | 0.39 | 0.94 | 0.09 | 2.41 | 1,5 |
| Elias 30 | K0III | 0.46 | 1.14 | 0.04 | 2.48 | 1,2 |
| HD 28225 | A3III | 0.12 | 0.34 | 0.02 | 2.83 | 2,3,6 |
| HD 28975 | A4III | 0.18 | 0.82 | 0.11 | 4.56 | 5,7,8 |
| HD 29333 | A2V | 0.20 | 1.03 | 0.07 | 5.15 | 5,9,12 |
| HD 29647 | B6-7IV | 0.37 | 0.64 | 0.05 | 1.73 | 5,7,10,11 |
| HD 29835 | K2III | 0.11 | 0.80 | 0.06 | 7.27 | 5,6,12 |
| HD 30168 | B8V | 0.14 | 0.59 | 0.06 | 4.21 | 5,6,8 |
| HD 30675 | B3V | 0.16 | 0.64 | 0.10 | 4.00 | 5,8,12 |
| HDE 279652 | A2V | 0.10 | 0.20[b] | 0.01 | 2.00 | 5,13 |
| HDE 279658 | B7V | 0.20 | 0.38[b] | 0.01 | 1.90 | 5,13 |
| HDE 283367 | B9V | 0.21 | 0.36 | 0.04 | 1.71 | 2,7,12 |
| HDE 283637 | A0V | 0.23 | 0.54[b] | 0.01 | 2.35 | 5,7,12 |
| HDE 283642 | A3V | 0.24[c] | 0.42 | 0.03 | 1.75 | 2,6 |
| HDE 283701 | B8III | 0.27 | 0.73 | 0.07 | 2.70 | 5,12 |
| HDE 283725 | F5III | 0.15 | 1.12 | 0.08 | 7.47 | 5,8,12 |
| HDE 283757 | A5V | 0.19[c] | 0.67 | 0.09 | 3.53 | 2,6 |
| HDE 283800 | B5V | 0.17 | 0.69[b] | 0.02 | 4.06 | 5,6 |
| HDE 283809 | B3V | 0.58 | 1.48 | 0.10 | 2.55 | 5,14 |
| HDE 283812 | A1V | 0.20 | 1.11 | 0.06 | 5.55 | 5,6,8,11,12 |
| HDE 283815 | A5V | 0.19[c] | 0.57 | 0.04 | 3.00 | 2,6 |
| HDE 283879 | F5V | 0.21[c] | 0.86 | 0.05 | 4.10 | 2,15 |
| RA 4 19 19.7 | — | 0.38 | 0.76 | 0.07 | 2.00 | 16 |
| RA 4 19 43.6 | — | 0.95 | 1.39 | 0.09 | 1.46 | 16 |
| RA 4 20 03.6 | — | 0.09 | 0.38 | 0.06 | 4.22 | 16 |
| RA 4 20 14.0 | — | 0.95 | 0.91 | 0.13 | 0.96 | 16 |
| RA 4 20 41.6 | — | 1.07 | 0.87 | 0.12 | 0.81 | 16 |
| RA 4 20 46.0 | — | 0.84 | 0.80 | 0.16 | 0.95 | 16 |
| RA 4 20 47.6 | — | 0.26 | 0.51 | 0.06 | 1.96 | 16 |
| RA 4 20 52.4 | — | 0.55 | 0.98 | 0.12 | 1.78 | 16 |
| RA 4 20 55.3 | — | 0.09 | 0.41 | 0.05 | 4.56 | 16 |

**Table 1.** continued

| Star I.D. | Sp Type[a] | $A_K$ mag | $p_K$ % | $\sigma_{p_K}$ % | $p_K/A_K$ % mag$^{-1}$ | Ref |
|---|---|---|---|---|---|---|
| RA 4 21 27.8 | — | 0.14 | 1.29 | 0.09 | 9.21 | 16 |
| RA 4 21 29.3 | — | 0.14 | 0.65 | 0.08 | 4.64 | 16 |
| RA 4 22 11.6 | — | 0.55 | 1.01 | 0.20 | 1.84 | 16 |
| RA 4 22 34.4 | — | 0.43 | 1.47 | 0.26 | 3.42 | 16 |
| Tam 8 | K5III | 2.33 | 2.7 | 0.3 | 1.16 | 4,17 |
| Tam 12 | — | 0.66 | 1.5 | 0.2 | 2.27 | 4 |
| Tam 16 | — | 0.60 | 2.0 | 0.2 | 3.33 | 4 |
| Tam 17 | — | 0.60 | 1.2 | 0.1 | 2.00 | 4 |

Notes – (a) If no spectral type is listed, type K5III has been assumed. In these cases, values of $A_K$ contain an error of about 0.2 mag due to uncertainty in assigned intrinsic color. (b) $p_K$ estimated using the values of $p_{\max}$, $\lambda_{\max}$ and $K$ from Whittet et al. (1992). Fractional error ($\sigma_{p_K}/p_K$) taken equal to that of $\sigma_{p_{\max}}/p_{\max}$. (c) $A_K = 0.56 \lambda_{\max} E_{B-V}$.

References– (1) Elias 1978; (2) Whittet et al. 1995; (3) Moneti et al. 1984; (4) Tamura et al. 1987; (5) Whittet et al. 1992; (6) Straižys & Meištas 1980; (7) Vrba & Rydgren 1985; (8) Kenyon et al. 1994; (9) Oja 1987; (10) Crutcher 1985; (11) Straižys et al. 1982; (12) Simbad database; (13) Ungerer et al. 1985; (14) Straižys 1985; (15) Slutskii et al. 1980; (16) Goodman et al. 1992; (17) Whittet et al. 1988.

**Figure 1.** Values of $p_K/A_K$ plotted against $A_K$ (from Table 1). Filled circles indicate the stars with known spectral types. Stars with uncertain spectral types are plotted using empty circles. The dotted line represents a least-squares fit to all data points: $p_K/A_K = 1.38 A_K^{-0.53}$.